\documentclass[prl,aps,showpacs,twocolumn]{revtex4}
\usepackage{graphicx}
\begin{document}

\title{Lasing in Single Cadmium Sulfide Nanowire Optical Cavities}

\author{Ritesh Agarwal, Carl J. Barrelet, and Charles M. Lieber}

\affiliation{Harvard University, Cambridge, Massachusetts 02138}

\date{\today}

\begin{abstract}

The mechanism of lasing in single cadmium sulfide (CdS) nanowire
cavities was elucidated by temperature-dependent and time-resolved
photoluminescence (PL) measurements. Temperature-dependent PL studies
reveal rich spectral features and show that an exciton-exciton
interaction is critical to lasing up to 75~K, while an exciton-phonon
process dominates at higher temperatures. These measurements together
with temperature and intensity dependent life-time and threshold
studies suggest that lasing is due to formation of excitons, and
moreover, have implications for the design of efficient, low-threshold
nanowire lasers.

\end{abstract}


\maketitle

Semiconductor nanowires (NWs) are emerging as versatile nanoscale
building blocks for the assembly of photonic devices
\cite{lieber03,wang01,duan01,duan03}, including polarization sensitive
photodetectors \cite{wang01}, light emitting diodes \cite{duan01}, and
electrical injection lasers \cite{duan03}.  Progress on
such nanophotonic devices will require developing
a detailed understanding of how confinement of charge carriers and
photons
affects optical properties and/or gives rise to interesting phenomena
\cite{duan03,johnson03,maslov03}. For example, single NWs have
recently been shown to function as optical waveguides
 and Fabry-Perot cavities
\cite{duan03,johnson02, johnson03, maslov03}. Intense optical excitation of single NWs has
produced stimulated emission and lasing
\cite{duan03,huang01,zapien04}, and significantly, lasing has also
been obtained from cadmium sulfide (CdS) NW electrical
injection devices \cite{duan03}. The mechanism of lasing in these
confined NW cavities is unclear, although critical to the rational development, for example, 
of ultralow threshold lasers.

\begin{figure}[t]
\begin{center}
\includegraphics[width=\columnwidth]{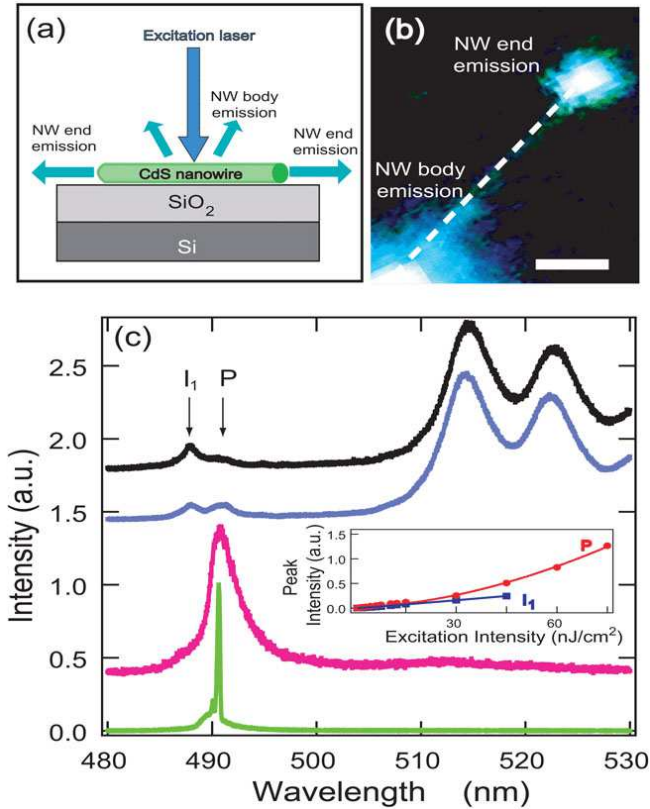}
\end{center}
\caption{
(a) Schematic of single NW optical experiments. (b) PL image showing luminescence 
from the excitation area (lower left) and one end (upper right) of a CdS NW. 
The NW was excited with a focused beam ($\sim 5$ $\mu$m diameter) with a power 
of 10 nJ/cm$^2$;  scale bar, 5 $\mu$m.  (c) PL spectra of CdS NW end emission 
recorded at 4.2 K with excitation powers of 0.6, 1.5, 30, and 240 nJ/cm$^2$ for 
the black, blue, red and green curves, respectively. Inset shows peak intensity 
of I$_1$ (black, squares) and P (red, circles) bands vs. incident laser power.  
Solid lines are fits to experimental data with power exponents of 0.95 for I$_1$ 
and 1.8 for P.}
\label{fig1}
\end{figure}

CdS nanowires are particularly interesting since both
optical excitation and electrical injection have been used to exceed
the threshold for lasing \cite{duan03}. In bulk CdS, three different
mechanisms are known to produce significant gain: exciton-exciton
scattering; exciton-logitudinal optic (ex-LO) phonon scattering; and
exciton-electron scattering
\cite{guillaume69,haug77,koch78,fischer74}.  The geometry and quality \cite{song84} of 
the bulk CdS materials has been found to affect 
the mechanism yielding the highest gain and
lasing, and importantly it is expected that geometrically confining cavities might also 
significantly affect the gain mechanism and lasing \cite{song84,ning01}.  
In this Letter we report detailed 
temperature-dependent and time-resolved PL measurements that define the mechanism of 
lasing in single nanowires for the first time. PL studies of CdS NW optical
cavities, which have diameters smaller than the wavelength of
light but large enough to sustain a single optical mode, demonstrate
that exciton-exciton scattering is critical to lasing up to 75~K and
that an exciton-phonon process dominates at higher temperatures. These
measurements together with intensity dependent PL, life-time, and
threshold studies imply that lasing is due to the formation of
excitons and not an electron-hole plasma at all of the temperatures
and intensities studied.

Single crystal CdS NWs with diameters from 80 to 150~nm and lengths up
to 100~$\mu$m were prepared by a metal nanocluster catalyzed
vapor-liquid-solid growth \cite{barrelet03}, and subsequently dispersed on  
Si/SiO$_2$ substrates
(600~nm thermal oxide) with an
average separation of 100~$\mu$m.
PL experiments were carried out by exciting the NWs
using frequency doubled Ti:sapphire laser pulses ($\sim$405~nm, 350~fs FWHM) 
and focused to a diameter of $\sim$75~$\mu$m.
 PL spectra from individual NWs were recorded
from the ends of the NW cavity as illustrated schematically in Fig.\ \ref{fig1}a 
using a homebuilt epi-fluorescence microscope (60$\times$, 0.7~NA objective)
with a spectral resolution ca.\ 0.15~nm. A typical PL image of
a NW excited by a tightly focused laser (Fig.\ \ref{fig1}b) exhibits
emission from the excitation region and 20~$\mu$m away at the NW end,
thereby demonstrating excellent waveguiding of the CdS NWs.  Below we focus
on the emission from the NW ends. 

\begin{figure}[htbp]
\begin{center}
\includegraphics[width=\columnwidth]{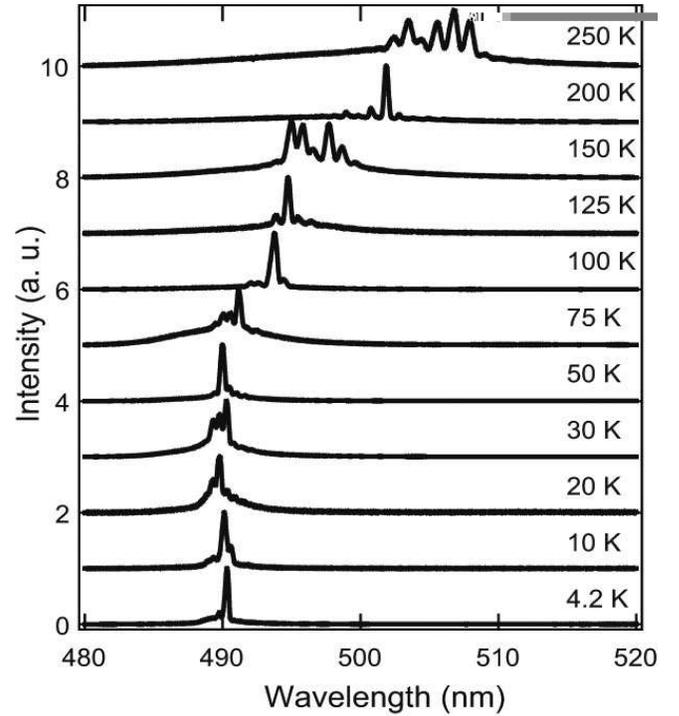}
\end{center}
\caption{
Temperature dependent spectra recorded from a single CdS NW above the threshold 
for lasing.  Temperature values are given next to each spectrum.  Excitation power 
ranges from 0.24 $\mu$J/cm$^2$ at 4.2 K to 3.3 $\mu$J/cm$^2$ at 250 K.  }
\label{fig2}
\end{figure}

PL spectra recorded at 4.2~K from a 40~$\mu$m long, ca.\ 100~nm
diameter CdS NW as a function of excitation power are shown in Fig.\
\ref{fig1}c; similar results were observed for NWs with 80-150~nm
diameters and 30-50~$\mu$m lengths. The PL spectrum recorded at low
excitation intensity of 0.6~nJ/cm$^2$ (Fig.\ \ref{fig1}c) reveals a
number of well-defined and reproducible features in the 488 to 530 nm
range, with peaks at 488.8, 490.5, 513, 522 and 530~nm. Comparison to
previous studies of bulk CdS crystals suggests that the peaks at 488.8
(I$_1$), 490.5~nm (P), 513 , and 522/530~nm correspond to neutral
acceptor bound excitons \cite{thomas62}, exciton-exciton scattering
\cite{magde70}, free electron-bound hole radiative recombination, and
the LO phonon progressions of the free electron-bound hole transitions
\cite{colbow66}, respectively.

Excitation power dependent measurements
(Fig. \ref{fig1}c) show that the intensity of the P-band increases rapidly and 
becomes dominant at higher excitation powers. Analysis of the P-band data shows 
an intensity increase of $\sim$ $I^{1.8}$, where $I$ is the excitation 
intensity (inset, Fig.\ \ref{fig1}c), while the I$_1$ band increases with 
an exponent near unity (0.95). These exponents are close to the values 
expected for 2- and 1-exciton processes \cite{guillaume69,magde70}. At 
higher excitation intensities the P-band is the dominant feature in PL 
spectra (Fig.\ \ref{fig1}c), and furthermore, shows a superlinear increase 
in the PL intensity above a threshold pumping power of 200~nJ/cm$^2$.  Above 
this threshold the PL spectrum collapses to a narrow peak at 490.5~nm with a 
line width of 0.3 nm indicative of lasing \cite{duan03,huang01}.  The position of 
the laser line and the power dependence of the PL spectrum strongly suggest that 
the mechanism of lasing in CdS NWs at 4.2~K is due to an exciton-exciton process. 
Although previous studies reported line narrowing at 490 nm \cite{duan03}, these 
studies did not identify the origin of the lasing peak.

\begin{figure}[htbp]
\begin{center}
\includegraphics[width=\columnwidth]{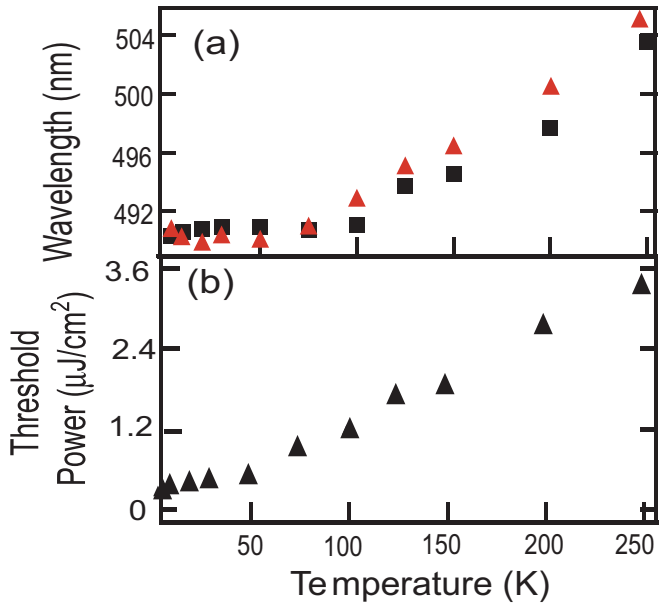}
\end{center}
\caption{
(a) Temperature dependence of key spectral features above and below threshold for lasing.
 Red triangles correspond to the temperature dependence of the most prominent lasing 
peak (for multiple peaks of comparable intensity, the central peak) vs. temperature. 
Black squares correspond to temperature dependence of the PL peaks that lead to lasing. 
These latter data were recorded at low excitation powers, from 25 nJ/cm$^2$ (4.2 K) to 
125 nJ/cm$^2$ (250 K). (b) Temperature dependence of the laser threshold power. }
\label{fig3}
\end{figure}

Temperature dependent data, which were recorded to probe further the mechanism of 
lasing in the CdS NWs, show that the lasing line at 490.5~nm is temperature independent 
from 4.2~K to 75~K, while the spectral features associated with lasing exhibit a 
pronounced red-shift at high temperatures (Fig.\ \ref{fig2}). A summary of these 
temperature dependent results (Fig. \ref{fig3}a) show clearly the temperature 
independence of the 490.5~nm peak above threshold at low temperatures, and the 
subsequent monotonic red-shift at 0.083~nm/K above ca.\ 100 K. Below the threshold, for lasing the main PL 
peak at 490.5~nm was also found to be temperature independent up to ca. 100~K.
Above 100~K, the PL peak shifts to ca. 494~nm and then further red-shifts at a rate similar to that observed above threshold. 
The peak at 494~nm has been assigned previously to exciton-LO 
band \cite{haug77,koch78,fischer74,song84}. The laser threshold temperature dependence
(Fig.\ \ref{fig3}b) shows that the threshold power increases very slowly 
until 50 K (4.5~nJ/cm$^2$/K) and then increases at a $3\times$ rate of 14~nJ/cm$^2$/K.

The observation of the dominant lasing line at 490.5~nm up to 75~K suggests that the 
exciton-exciton scattering process is the predominant gain mechanism for CdS NW optical 
cavities in this lower temperature regime. This temperature independence is consistent 
with the ca. 20~meV separation between the exciton $n$ = 1 ground and $n$ = 2 excited state 
\cite{koch78}. Hence, $n$ = 2 and higher exciton states are not significantly populated  
over this temperature range, and the dominant lasing line should not 
vary strongly with temperature \cite{haug77,koch78,fischer74} as we observe. In addition,
measurements of the threshold power dependence for our CdS NW system (Fig.\ \ref{fig3}b) 
show little temperature dependence below 75~K, which is also consistent with the 
interpretation that lasing arises primarily from $n$ = 1 excitons.

At temperatures above 75~K, higher exciton states become thermally populated and 
will lead to increases in the lasing threshold for the exciton-exciton process.  
Exciton scattering by LO phonons can contribute more to
gain at these temperatures since the LO phonon energy, 38~meV, is larger than the available 
thermal energy
\cite{haug77,koch78,fischer74}.  Significantly at temperatures higher than 75~K 
the PL band 
and laser lines occurs in the region of the 
ex-LO phonon scattering band, 493~ nm \cite{koch78,fischer74}.  
Temperature-dependent measurements (Fig.\ \ref{fig3}a) further 
show a red-shift, 0.083~nm/K,
consistent with that observed in bulk CdS \cite{fischer74}, 
0.086~nm/K, for the ex-LO phonon band. 
Therefore, at temperatures higher than 75~K 
the lasing mechanism of CdS NWs can be assigned to ex-LO phonon scattering process. 

In contrast, lasing in bulk CdS crystals has been been primarily 
attributed at higher temperature to 
an exciton-electron scattering process \cite{guillaume69,haug77,koch78,fischer74}.  
The exciton-electron scattering band appears around 494 nm at 70 K and 
exhibits a large red-shift of ca. 0.147~nm/K \cite{koch78,fischer74}. 
The temperature dependent 
red shift  we find for CdS NWs, 0.083~nm/K,(Fig.\ \ref{fig3}a) is much smaller
than this but agrees with the ex-LO process (see above). We believe that the 
difference between previous bulk studies and these new NW lasers arises from the small 
modal volume of the laser cavity \cite{volume}. To understand this unique difference a 
simple model based on the density of states of electrons and excitons at 
equilibrium \cite{agarwal01} was used
to calculate the ratio of free carriers to the total injected carriers 
in the cavity as a function of temperature. These results show that at 150~K
and low carrier densities (10$^{15}$ - 10$^{16}$ ~cm$^{-3}$), 
$\sim$80\% of the carriers exist as free 
electrons, whereas at higher densities (10$^{18}$cm$^{-3}$), 
the free carriers do not exceed $\sim$40\%. The small volume of 
the NW laser cavity can effectively yield large carrier densities, and hence the portion 
of free carriers is much smaller than in bulk samples thus reducing 
contributions from exciton-electron scattering.

\begin{figure}[htbp]
\begin{center}
\includegraphics[width=\columnwidth]{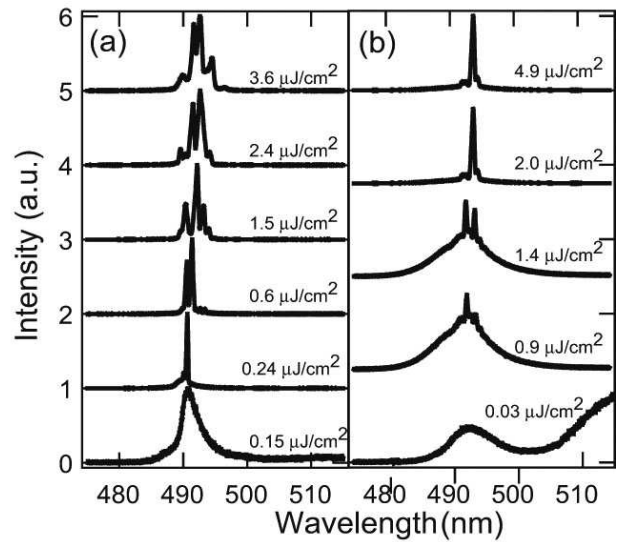}
\end{center}
\caption{
Spectra recorded above the lasing threshold for a CdS NW at different pumping powers 
at (a) 4.2 K and (b) 100 K. The bottom spectrum in both the panels is the PL recorded 
below the threshold for lasing.}
\label{fig4}
\end{figure}

In addition, we find that the laser threshold power for CdS NWs is ca. 10 times less sensitive
to temperature than reported for
bulk CdS crystals \cite{hvam71,leheney70}.  The linear increase of the lasing 
threshold power with temperature due to ex-LO mechanism of lasing cannot be explained 
by a simple model based on the temperature dependence of the occupation of the phonon 
states as this model predicts an exponential increase in the threshold \cite{haug77,koch78}. 
We believe 
this disagreement can be attributed to the strong confinement of the optical mode 
in the CdS NW cavity, where there is $\sim$50\% mode confinement due to strong 
dielectric mismatch between the wire and its surroundings \cite{modeconfinement}. 
Strong mode confinement in optical cavities leads to much better overlap of the 
optical mode with the excitons, where the modal volume of the NW cavity is of the 
order of $3\lambda^3$ \cite{volume}. This observation is similar to the reduced magnitude
and lower temperature-dependence of the laser threshold observed in double heterostructure
planar semiconductor lasers \cite{panish70}, although the mode confinement is typically 
only a few percent. Further theoretical analysis which includes the effect of strong 
optical confinement \cite{ning01} along with the exciton-LO scattering mechanism for 
gain will be required to explain the weak temperature dependence of lasing threshold 
observed for CdS NWs.

We have also investigated the possibility of lasing from an electron-hole plasma, 
which can form due to screening of excitons at high carrier densities 
($n \sim 2.1 \times 10^{18}$cm$^{-3}$) \cite{rice77}, by recording 
spectra as a function of pumping powers above the threshold for lasing. Significantly, 
no red-shift of the lasing line was observed at either 4.2~K or 100~K while increasing the 
excitation intensity approximately an order of magnitude (Fig.\ \ref{fig4}).
An estimate of the carrier density based on 
laser pump intensity (5~$\mu$J/cm$^2$), CdS NW dimensions 
(40~$\mu$m length, 100~nm diameter), CdS absorbance at 400~nm ($\sim10^5$~cm$^{-1}$) 
gives a value of $n \sim1.0 \times 10^{18}$~cm$^{-3}$, which is close to the lower 
threshold reported \cite{rice77} to yield dissociation into an electron-hole plasma.

Formation of electron-hole plasma is generally accompanied by a strong red shift 
of the spectrum owing to band-gap renormalization effect \cite{fischer74,koch93}, 
and has been suggested for GaN nanowires \cite{johnson02}. The absence of 
red shift of lasing lines with increasing excitation intensity is a further proof of 
lasing by excitonic mechanism in the range of the excitation intensities used in our 
study.  In addition, excited state PL life-time measurements 
(data not shown) obtained on CdS NWs did not show significant pump intensity dependence.  
The lifetime at 4.2~K was observed to be 1.2~ns (90\% amplitude) which rapidly dropped 
to 400~ps at 50~K and then decayed very slowly to $\sim$350 ps at room temperature.  
However, neither the amplitude nor the timescale of the PL lifetime changed more than 
10\% with $10\times$ - $5\times$ (4.2~K - 100~K) increase in pump intensity.  In contrast, 
previous studies of highly excited CdS showed a rapid 200~ps (5~K) lifetime that was 
attributed to the decay of 
an electron-hole plasma to excitons \cite{saito85}. The absence of a fast component 
in our experiments even at very high pump powers 
further rule out the formation of electron-hole plasma in the CdS NW lasers.

In summary, the mechanism of lasing in single CdS NWs was elucidated by temperature 
dependent spectroscopic studies.  The data from these studies show that the mechanism
 is exciton-based: exciton-exciton scattering from 4.2 - 75~K and exciton-LO scattering 
at higher temperatures.  Similar detailed optical studies would be useful for other NW 
optical cavities, such as ZnO \cite{johnson03} and GaN \cite{johnson02}, to enable 
determination of the lasing mechanism, since the mechanism will depend upon factors 
including the crystal properties, cavity configuration, losses, and excited state 
lifetimes. We believe our results will help in modeling \cite{maslov03,ning01} the lasing 
behavior in these novel nanostructures, and also aid in the design of ultralow threshold
 NW lasing devices. For example, the excitonic mechanism of lasing 
could be enhanced by using radial NW heterostructures \cite{lieber03}. 
Specifically, small diameter NWs coated by a larger bandgap material might be used as an 
efficient nanoscale lasing structures, where the small diameter NW provides a low 
threshold 
active medium due to exciton confinement, and the outer coating would facilitate 
waveguiding in the cavity.


\begin{thebibliography}{99}

\bibitem{lieber03}  C. M. Lieber, MRS Bull. {\bf28}, 486 (2003).

\bibitem{wang01}  J. F. Wang et al., Science {\bf293}, 1455 (2001).

\bibitem{duan01}  X. F. Duan et al., Nature {\bf409}, 66 (2001).

\bibitem{duan03}  X. F. Duan et al., Nature {\bf421}, 241 (2003).

\bibitem{johnson03}  J. C. Johnson et al., J. Phys. Chem. B {\bf107}, 1186 (2003).

\bibitem{maslov03}  A. V. Maslov and C. Z. Ning, Appl. Phys. Lett. {\bf83}, 1237 (2003).


\bibitem{johnson02}  J. C. Johnson et al., Nat. Mat. {\bf1}, 106 (2002).

\bibitem{huang01}  M. H. Huang et al., Science {\bf292}, 1897 (2001).

\bibitem{zapien04}  J. A. Zapien et al, Appl. Phys. Lett. {\bf84}, 1189 (2004).

\bibitem{guillaume69}  C. B. A La Guillaume, J.-M. Debever, and F. Salvan, Phys. Rev. {\bf177}, 567 (1969).

\bibitem{haug77}  H. Haug and S. W. Koch, Phys. Stat. Sol. B {\bf82}, 531 (1977).

\bibitem{koch78}  S. W. Koch et al., Phys. Stat. Sol. B {\bf89}, 431 (1978).

\bibitem{fischer74}  T. Fischer and J. Bille, J. Appl. Phys. {\bf45}, 3937 (1974).

\bibitem{song84}  J. J. Song and W. C. Wang, J. Appl. Phys. {\bf55}, 660 (1984).

\bibitem{ning01}  A. V. Maslov and C. Z. Ning, IEEE J. of Quant. Electron. {\bf40}, 1389 (2004).

\bibitem{barrelet03}  C. J. Barrelet et al., J. Am. Chem. Soc. {\bf125}, 11498 (2003).

\bibitem{thomas62}  D. G. Thomas and J. J. Hopfield, Phys. Rev. {\bf128}, 2135 (1962).

\bibitem{magde70}  D. Magde and H. Mahr, Phys. Rev. Lett. {\bf24}, 890 (1970).

\bibitem{colbow66}  K. Colbow, Phys. Rev. {\bf141}, 742 (1966).

\bibitem{volume}  For CdS NW, $d$ = 100 nm; $L$ = 40 $\mu$m, $\lambda$ = 500 nm; therefore, the volume occupied by the CdS material assuming a cylindrical geometry is $\sim$ $3\lambda^3$.

\bibitem{agarwal01}  R. Agarwal, C. J. Barrelet, and C. M. Lieber, unpublished results.

\bibitem{hvam71}  J. M. Hvam, Phys. Rev. B. {\bf4}, 4459 (1971).

\bibitem{leheney70}  R. F. Leheney et al., Appl. Phys. Lett. {\bf17}, 494 (1970).

\bibitem{modeconfinement} Mode confinement in a cylindrical waveguide is given by $\eta = 1 - \{2.405\exp(-1/V)\}^2/V^{-3}$; $V = kr(n^2 - 1)^{1/2}$ and $k = 2\pi/\lambda$;
 For CdS, $n \sim 2.6$; $\lambda$ = 500 nm.

\bibitem{panish70}  M. B. Panish et al., Appl. Phys. Lett. {\bf16}, 326 (1970).

\bibitem{rice77}  T. M. Rice, Solid State Phys. {\bf32}, 1 (1977).

\bibitem{koch93}  S. W. Koch, Quantum Theory of the Optical and Electronic Properties of Semiconductors (World Scientific, Singapore, 1993).

\bibitem{saito85}  H. Saito and E. O. Gobel, Phys. Rev. B. {\bf31}, 2360 (1985).




\end{thebibliography}
\end{document}